\documentstyle[12pt]{article}

\def\C{{\rm\kern.24em \vrule width.02em height1.4ex depth-.05ex \kern-.26em
C}}

\newcommand{\beq}{\begin{equation}}
\newcommand{\eeq}{\end{equation}}

\begin{document}
\baselineskip=20pt
\pagestyle{plain}

\title{Liouville Theory: Ward Identities for Generating Functional and
Modular Geometry}

\author{
Leon A.~Takhtajan \\
Department of Mathematics\\
SUNY at Stony Brook \\
Stony Brook, NY 1794-3651\\
U.S.A.
}
\date{}
\maketitle

\begin{center}
{\bf Abstract}
\end{center}
\begin{quote}
We continue the study of quantum Liouville theory through Polyakov's functional
integral \cite{Pol1,Pol2}, started in \cite{T1}. We derive the perturbation
expansion for Schwinger's generating functional for connected multi-point
correlation functions involving stress-energy tensor, give the ``dynamical''
proof of the Virasoro symmetry of the theory and compute the value of the
central charge, confirming previous calculation in \cite{T1}. We show that
conformal Ward identities for these correlation functions contain such basic
facts from K\"{a}hler geometry of moduli spaces of Riemann surfaces, as
relation between accessory parameters for the Fuchsian uniformization,
Liouville action and Eichler integrals, K\"{a}hler potential for the
Weil-Petersson metric, and local index theorem. These results affirm the
fundamental role, that universal Ward identities for the generating functional
play in Friedan-Shenker modular geometry \cite{FS}.

\end{quote}

{\bf 1} According to \cite{Pol1,Pol2,T1}, the correlation function of puncture
operators in Liouville theory is given by the following functional integral
\beq \label{FI}
<X>={\cal\int}_{{\cal C}(X)} {\cal D}\phi~e^{-(1/2 \pi h)S(\phi)},
\eeq
where $X=\hat{\C} \setminus \{w_1, \ldots, w_n\},~w_i \neq w_j~{\rm for}~i \neq
j$, is the $n$-punctured Riemann sphere. ``Domain of integration''
${\cal C}(X)$ consists of conformal metrics $ds^2= e^{\phi(w, \bar{w})}
|dw|^{2}$ on $X$, satisfying the following asymptotics near punctures
$w_1, \ldots, w_{n-1}, w_n=\infty$,
$$e^{\phi} \cong \frac{1}{r^{2}_{i} \log^{2}r_{i}},~i=1, \dots ,n,$$
where $r_{i}=|w-w_{i}|,\ i=1, \ldots ,n-1,~r_n=|w|$, and $w$ is the
global complex coordinate on $X$. Also $h>0$ is a coupling constant and
functional $S(\phi)$ is Liouville action \cite{ZT},
$$ S(\phi)= \lim_{\epsilon \rightarrow 0} \{  \int_{X_{\epsilon}}
(|\phi_{w}|^{2} +e^{\phi})d^{2}w +2\pi n \log\epsilon +4\pi (n-2)\log|
\log\epsilon| \}, $$
where $X_{\epsilon}= X \setminus  \bigcup^{n-1}_{i=1}\{|w-w_{i}|<\epsilon\}
\bigcup \{|w|>1/\epsilon\}$. Classical equations of motion $\delta S=0$
yield Liouville equation---the equation for complete
conformal metric on $X$ of constant negative curvature $-1$. It has
a unique solution, called Poincar\'{e}, or hyperbolic metric, and is denoted by
$\phi_{cl}$.

As in \cite{T1}, we define $<X>$---the expectation value of the Riemann
surface $X$---by its perturbation expansion around the classical solution
$\phi= \phi_{cl}$ (thus making a choice of the ``integration measure'' in
(\ref{FI})). Corresponding propagator $G(w,w^{\prime})$ is given by the
Green's function of operator $2\Delta +1$, where
$$\Delta=-e^{-\phi_{cl}} \partial^{2}_{w \bar{w}}$$
is a hyperbolic Laplacian on $X$. Logarithmic divergence of $G$ at coincident
points is renormalized in a reparametrization invariant way using the geodesic
distance in the Poincar\'{e} metric (see. e.g., \cite{T1}).

{\bf 2} Conformal invariance of Liouville theory implies, in particular, that
its stress-energy tensor is traceless. Namely, its $(2,0)$-component
$T(\phi)(w)$ is given by
$$T(\phi)=\frac{1}{h}(\phi_{ww}-\frac{1}{2}\phi_{w}^{2}),$$
is conserved on classical equations of motion
$$\partial_{\bar{w}} T_{cl}=0,$$
and has the transformation law of the projective connection (times $1/h$)
under holomorphic change of coordinates, i.e.
$$\tilde{T}(\tilde{w})=T(f(\tilde{w}))(f^{\prime}(\tilde{w}))^{2}+
\frac{1}{h}{\cal S}(f)(\tilde{w}),~~w=f(\tilde{w}).$$
Here ${\cal S}$ stands for the Schwarzian derivative,
$${\cal S}(f)=\frac{f^{\prime \prime \prime}}{f^{\prime}}-
\frac{3}{2} (\frac{f^{\prime \prime}}{f^{\prime}})^{2}.$$
Similarly, the $(0,2)$-component $\bar{T}(\phi)(w)$ of the stress-energy tensor
is given by
$$\bar{T}(\phi)=\frac{1}{h}(\phi_{\bar{w}\bar{w}}-\frac{1}{2}\phi_{\bar{w}}^{2}
).$$

According to Belavin-Polyakov-Zamolodchikov \cite{BPZ}, conformal
symmetry manifests itself through the infinite sequence of conformal Ward
identities, that relate correlation functions involving the stress-energy
tensor components with correlation functions without them. Specifically, we
define multi-point expectation values of the holomorphic and anti-holomorphic
components of the stress-energy tensor as
$$<\prod_{i=1}^{k}T(u_i) \prod_{j=1}^{l}\bar{T}(\bar{v_j})X>=
{\cal \int}_{{\cal C}(X)} {\cal D}\phi~\prod_{i=1}^{k}T(\phi)(u_i)\prod_{j=1}^{
l}\bar{T}(\phi)(\bar{v_j})~ e^{(-1/2\pi h) S(\phi)},$$
and denote by $<<\cdots>>$ their normalized connected forms. Thus, for
example,
\begin{eqnarray*}
<<T(w)X>> &=& \frac{<T(w)X>}{<X>},~ ~<<\bar{T}(\bar{w})X>>~=~
\frac{<\bar{T}(\bar{w})X>}{<X>}, \\
<<T(u)T(v)X>> &=& \frac{<T(u)T(v)X>}{<X>}~-~<<T(u)X>><<T(v)X>>,\\
<<T(u)\bar{T}(\bar{v})X>> &=& \frac{<T(u)\bar{T}(\bar{v})X>}{<X>}~-~<<T(u)X>>
<<\bar{T}(\bar{v})X>>,~~{\rm etc.}
\end{eqnarray*}
These correlation functions should satisfy conformal Ward identities of BPZ
\cite[formulas (3.10), (3.15)]{BPZ}, that we present in the following
succinct form (after fixing overall ${\rm SL}(2, \C)$-symmetry by normalizing
$w_{n-2}=0, w_{n-1}=1, w_n=\infty$, and using \cite[formulas (A.7)]{BPZ}).

Namely, denote by $\Delta(h)=\bar{\Delta}(h)$ conformal weights of the
puncture operator, by $c$---the central charge of the Virasoro algebra, and
by ${\cal L}(w)$ and $\bar{{\cal L}}(\bar{w})$---the following first order
differential operators
$${\cal L}(w)=\sum_{i=1}^{n-3}R(w,w_i)\partial_{w_i},~
\bar{{\cal L}}(\bar{w})=\sum_{i=1}^{n-3}R(\bar{w},\bar{w}_i)
\partial_{\bar{w}_i},$$
where
$$R(w,w_i)=\frac{1}{w-w_i}+\frac{w_i-1}{w}-\frac{w_i}{w-1}=\frac{w_i(w_i-1)}
{(w-w_i)w(w-1)}.$$
We have
\beq \label{T}
<<T(w)X>>_0 ~\doteq~<<T(w)X>>-~T_{s}(w)={\cal L}(w) \log<X>,
\eeq
where
$$T_{s}(w)=\sum_{l=1}^{n-1}\frac{\Delta(h)}{(w-w_l)^2}+\frac{(2-n)\Delta(h)}
{w(w-1)},$$
and analogous expression for $<<\bar{T}(\bar{w})X>>$. The Ward
identities for two-point correlation functions have the form
\beq \label{TT}
<<T(u)T(v)>>=\frac{c/2}{(u-v)^4}+\{2R_{v}(u,v)+R(u,v)\partial_v +{\cal L}(u) \}
<<T(v)X>>,
\eeq
\beq \label{TaT}
<<T(u)\bar{T}(\bar{v})X>>={\cal L}(u)<<\bar{T}(\bar{v})X>>=
{\cal L}(u)\bar{{\cal L}}(\bar{v})\log<X>.
\eeq
Similar formulas can be obtained for the Ward identities for multi-point
correlation functions (see Sect.~4). According to BPZ \cite{BPZ}, relations
(\ref{T})---(\ref{TT}) state, at the level of correlation functions, that
puncture operators are primary fields and $T(w), \bar{T}(\bar{w})$ are
generating functions of the holomorphic and anti-holomorphic Virasoro algebras,
that mutually commute in virtue of (\ref{TaT}).

{\bf 3} Since we have defined expectation values through functional integrals,
we need to affirm the validity of conformal Ward identities (\ref{T})---(\ref
{TaT}).  This will be done by using the perturbation theory.

We start with formula (\ref{T}), which, according to \cite{T1}, encodes
important information about modular geometry---a K\"{a}hler geometry of
the modular space, in our case---moduli space ${\cal M}_n$ of Riemann
surfaces of genus $0$ with $n > 3$ punctures. Validity of (\ref{T}) at the tree
level is equivalent to the relation between accessory parameters of the
Fuchsian uniformization and the classical Liouville action, conjectured in
\cite{Pol2} and proved in \cite{ZT} (see \cite{T} for review). In the one-loop
approximation, validity of (\ref{T}) yields an explicit formula for the first
derivative of the Selberg zeta function (evaluated at the special point $s=2$)
with respect to the moduli parameters proved in \cite{ZT2}.

It is remarkable, that Ward identity (\ref{T}) fits perfectly well into the
general philosophy of  Friedan-Shenker modular geometry \cite{FS}, that
interpets the expectation value $<X>$ as a Hermitian metric in a certain
holomorphic line bundle over ${\cal M}_n$, and quadratic differential $<<T(w)X>
>_{0}dw^2$---as a $(1,0)$-component of the canonical metric connection. Indeed,
according to \cite[Section 2.6]{ZT}, quadratic differentials $-R(w,w_i)dw^2$,
considered as a $(1,0)$-forms on ${\cal M}_n$, are dual to Beltrami
differentials, representing vector fields $\pi \partial_ {w_i},~i=1, \ldots,
n-3,$ on ${\cal M}_n$. Thus, ${\cal L}(w)$ can be interpreted as
$(1,0)$-component $\partial$ (times $-1/\pi$) of exterior differential $d=
\partial + \bar{\partial}$ on ${\cal M}_n$. Therefore, formula (\ref{T}) reads
\beq \label{TF}
<<T(w)X>>_{0}dw^2=-\frac{1}{\pi}\partial \log<X>,
\eeq
which should be compared with \cite[formula (6)]{FS}.

Next, consider connected two-point correlation function $<<T(u)T(v)X>>$.
At the tree level, we get
\beq \label{TTT}
<<T(u)T(v)X>>_{tree}=\frac{2\pi}{h}(\partial_{u}^{2}-(\phi_{cl})_u
\partial_u) (\partial_{v}^2-(\phi_{cl})_v \partial_v)G(u,v).
\eeq
Expression (\ref{TTT}) has $(6/h)(u-v)^{-4}$ as leading singularity at the
diagonal $u=v$, affirming that $c_{cl}=12/h$. Substituting (\ref{TTT}) into
the Ward identity (\ref{TT}) and using $<<T(v)X>>_{tree}= T_{cl}(u)$, we see
that validity of (\ref{TT}) at the tree level is equivalent to the relation
between derivatives of accessory parameters and Eichler integrals (see, e.g.,
\cite[Ch.~5]{Kra}) proved in \cite{ZT}. At the one-loop level, we get
$2 \pi^2 G_{uv}^{2}(u,v)$ as the leading singular term in $<<T(u)T(v)X>>$,
which produces the quantum correction $c_{loop}=1$ to the central charge of the
Virasoro algebra. Inspection of higher loops shows that they do not
contain singular terms of order $(u-v)^{-4}$. Thus, $c_{Liouv}=1+12/h$, as
was calculated in \cite{T1}. Moreover, validity of the Ward identity
(\ref{TT}) in all loops is equivalent to remarkable (and rather complicated!)
relations between Green's functions and their derivatives with respect to the
moduli parameters, which can be verified directly. This provides a ``dynamical
proof'' of the Virasoro algebra symmetry of the Liouville theory.

It is also instructive to examine the two-point correlation function
$<<T(u)\bar{T}(\bar{v})X>>$. Corresponding Ward identity (\ref{TaT}) can
be rewritten in the form
\beq \label{K}
<<T(u)\bar{T}(\bar{v})X>>du^2d\bar{v}^{2}=\frac{1}{\pi^2}\partial
\bar{\partial} \log<X>,
\eeq
which allows to interpret $<<T(u)\bar{T}(\bar{v})X>>$ as a $(1,1)$-component
of the curvature form for the connection in the holomorphic line bundle over
${\cal M}_n$, associated with Hermitian metric $<X>$ (cf.~\cite[formula
(15)]{FS}).
At the tree level
\beq \label{WP}
<<T(u)\bar{T}(\bar{v})X>>_{tree}=\frac{2 \pi}{h}(\partial_{u}^{2}-(\phi_{cl})_u
\partial_u)(\partial_{\bar{v}}^{2}-(\phi_{cl})_{\bar{v}}\partial_{\bar{v}})
G(u,v).
\eeq
This expression can be shown to coincide with Hermitian form of the
Weil-Petersson metric on moduli space ${\cal M}_{n}$. Since $\log<X>_{tree}
=-(1/2 \pi h) S_{cl}$, we see that at the tree level, the Ward identity
(\ref{K}) yields the classical Liouville action as a (local) K\"{a}hler
potential of the Weil-Petersson metric on ${\cal M}_n$! This fact was proved
in \cite{ZT}. Similarly, validity of (\ref{K}) in the one-loop approximation
is equivalent to the local index theorem for the families of punctured Riemann
surfaces \cite{ZT2}.

{\bf 4} Examined cases show that conformal Ward identities encode important
information about K\"{a}hler geometry of moduli spaces. Multi-point correlation
functions provide further insight, affirming the profound role of Ward
identities in modular geometry. In this respect we derive perturbative
expansion for the generating functional and present the universal Ward
identities it satisfies.

Following Schwinger \cite{JS}, we consider a bounded Beltrami differential
$\mu d\bar{w}/dw$ on Riemann surface $X$ as an external source and define
the generating functional for normalized multi-point correlation
functions of the stress-energy tensor by the following expression
\beq \label{z}
{\cal Z}(\mu,\bar{\mu};X)=\frac{Z(\mu,\bar{\mu};X)}{<X>},
\eeq
and
\beq \label{Z}
Z(\mu,\bar{\mu};X)={\cal \int}_{{\cal C}(X)}{\cal D}\phi \exp\{-\frac{1}
{2 \pi h}S(\phi) + {\rm v.p.}\int_{X}(T(\phi)\mu + \bar{T}(\phi)\bar{\mu})
d^{2}w\}.
\eeq
Here, since Riemann surface $X$ has a global coordinate $w$, we can consider
$(2,0)$-component of the stress-energy tensor ($(0,2)$-component) as quadratic
differential $Tdw^2$ ($\bar{T}d\bar{w}^2$), so that $T\mu dw \wedge d \bar{w}$
($\bar{T}\bar{\mu} dw \wedge d\bar{w}$) is a $(1,1)$-form on $X$. In general
case, one should choose a projective connection $T_0$ and use quadratic
differential $(T-T_0)dw^2$ (cf.~the difference $<<T(w)X>>-~T_{s}(w)$ in
(\ref{T})). The integral in (\ref{Z}) is the principal value integral in
virtue of the second order poles of $T$ and $\bar{T}$ at the punctures.

Generating functional for the connected multi-point correlation functions
is defined as
\beq \label{W}
\frac{1}{h} {\cal W}(\mu,\bar{\mu};X)=\log {\cal Z}(\mu, \bar{\mu};X),
\eeq
so that
$$h<<\prod_{i=1}^{k}T(u_i) \prod_{j=1}^{l}\bar{T}(\bar{v}_j)X>>=
\frac{\delta^{k+l}{\cal W}(\mu,\bar{\mu};X)}{\delta \mu(u_1) \cdots \delta \mu
(u_k)\delta \bar{\mu}(\bar{v}_1) \cdots \delta \bar{\mu}(\bar{v}_l)}|_{\mu=\bar
{\mu}=0}.$$

Expanding around the classical solution $\phi=\phi_{cl}$, we derive the
perturbation expansion for the functional integral (\ref{Z}). The final result
reads
\begin{eqnarray*}
Z(\mu, \bar{\mu};X) &=& \exp\{{\rm v.p.}\int_{X}(T_{cl}\mu+\bar{T}_{cl}
\bar{\mu})d^{2}w- \frac{1}{2\pi h}S_{int}(h \frac{\delta}{\delta \chi}) \}  \\
& &\mbox{} \times \exp\{\frac{\pi}{h}\int_{X}\chi G(1-2 \pi Ge^{-\phi_{cl}}
\partial_w \mu \partial_w-2 \pi Ge^{-\phi_{cl}}\partial_{\bar{w}}\bar{\mu}
\partial_ {\bar{w}})^{-1} (\chi) d^{2}w \}_{|_{\chi=f+\bar{f}}} \\
& & \mbox{} \times {\rm Det} (1-2 \pi Ge^{-\phi_{cl}}\partial_w \mu \partial_w-
2 \pi Ge^{-\phi_{cl}}\partial_{\bar{w}} \bar{\mu}\partial_{\bar{w}})^{-1/2}.
\end{eqnarray*}
where $G=(2\Delta +1)^{-1},~f=e^{-\phi_{cl}}(\mu_{ww}+ ((\phi_{cl})_w \mu)_w)$
is a globally defined function on $X$,
and
$$S_{int}(\psi)=\sum_{k=3}^{\infty}\frac{1}{k!}\int_{X}\psi^k e^{\phi_{cl}}
d^{2}w.$$
In order to obtain perturbation expansion for the generating functional
${\cal W}$, one should use the formula
$$\log {\rm Det}(1-A)=\sum_{k=1}^{\infty}\frac{1}{k}{\rm Tr}A^k,$$
and the reparametrization invariant regularization procedure. Previously
considered examples are special cases of this scheme.

Conformal Ward identities (\ref{T})---(\ref{TaT}) (or, more precisely,
corresponding operator product expansions) are equivalent to the universal
Ward identities for the generating functional ${\cal W}$. They formally
coincide with the Ward identity in the light-cone gauge, derived by Polyakov
\cite{Pol3} (see also \cite[formulas (10)-(12)]{FS} and (\cite[formula
(2.9)]{V}). Namely, we have
\begin{eqnarray*}
(\partial_{\bar{w}}+\pi \mu \partial_{w}+2 \pi \mu_w)\frac{\delta
{\cal W}} {\delta \mu(w)}(\mu,\bar{\mu};X)\\
=-\frac{\pi hc}{12} \partial_{w}^{3}\mu + h \partial_{\bar{w}}T_{s}(w) +
\partial_{\bar{w}} {\cal L}(w) \{{\cal W}(\mu, \bar{\mu};X)+h \log<X>\},
\end{eqnarray*}
and
\begin{eqnarray*}
(\partial_w + \pi \bar{\mu} \partial_{\bar{w}} +2 \pi \bar{\mu}_{\bar{w}})
\frac{\delta{\cal W}}{\delta \bar{\mu}(w)}(\mu,\bar{\mu};X)\\
=-\frac{\pi hc}{12} \partial_{\bar{w}}^{3}\bar{\mu} + h \partial_w \bar{T}_{s}
(\bar{w}) + \partial_w \bar{{\cal L}}(\bar{w})\{{\cal W}(\mu,\bar{\mu};X)+
h\log<X>\}.
\end{eqnarray*}

Using the formula
$$\partial_{\bar{u}}R(u,v)=\pi \delta(u-v),$$
where $u,v \neq 0,1$, one immediately gets (\ref{T})---(\ref{TaT}) from the
universal Ward identities. We anticipate their fundamental role in the modular
geometry, which will be discussed elsewhere.

We finally note, that our results can be generalized for compact Riemann
surfaces of genus $>1$. In this case, in order to formulate Ward identities
and to define the generating functional for correlation functions, one needs
to choose a projective connection $T_0$ and to consider the difference $T-T_0$
(cf.~\cite{S}).

{\bf Acknowledgments.} I appreciate stimulating conversations with A.~Polyakov.
This research was supported in part by the NSF grant DMS-92-04092.

\end{document}